\documentclass[12pt,preprint]{emulateapj} 
\usepackage{graphicx}
\usepackage{epstopdf}

\newcommand{\afx}{PS1-10afx}

\newcommand{\rps}{\ensuremath{r_{\rm P1}}}
\newcommand{\ips}{\ensuremath{i_{\rm P1}}}
\newcommand{\zps}{\ensuremath{z_{\rm P1}}}
\newcommand{\yps}{\ensuremath{y_{\rm P1}}}

\newcommand{\simlt}{\mathrel{\hbox{\rlap{\hbox{\lower4pt\hbox{$\sim$}}}\hbox{$<$}}}}
\newcommand{\simgt}{\mathrel{\hbox{\rlap{\hbox{\lower4pt\hbox{$\sim$}}}\hbox{$>$}}}}
\newcommand{\Dls}{{D_{\rm ls}}}
\newcommand{\Ds}{{D_{\rm s}}}
\newcommand{\Dl}{{D_{\rm l}}}

\begin{document}

\title[A Magnified SNIa]{Extraordinary Magnification of the Ordinary Type\,Ia Supernova PS1-10afx}

\author{
  Robert M. Quimby\altaffilmark{1,2},
  Marcus C. Werner\altaffilmark{1},
  Masamune Oguri\altaffilmark{1},
  Surhud More\altaffilmark{1},
  Anupreeta More\altaffilmark{1},
  Masayuki Tanaka\altaffilmark{1},
  Ken'ichi Nomoto\altaffilmark{1},
  Takashi J. Moriya\altaffilmark{1},
  Gaston Folatelli\altaffilmark{1},
  Keiichi Maeda\altaffilmark{1}
  \&
  Melina Bersten\altaffilmark{1}
}

\altaffiltext{1}{
  Kavli Institute for the Physics and Mathematics of the Universe (Kavli IPMU), 
  The University of Tokyo, 
  5-1-5 Kashiwanoha, 
  Kashiwa-shi, Chiba, 277-8583, Japan 
}
\altaffiltext{2}{
  email: robert.quimby@ipmu.jp
}

\begin{abstract}

Recently, Chornock and co-workers announced the Pan-STARRS discovery
of a transient source reaching an apparent peak luminosity of $\sim$$4
\times 10^{44}$\,erg\,s$^{-1}$. We show that the spectra of this
transient source are well fit by normal Type\,Ia supernova (SNIa)
templates. The multi-band colors and light-curve shapes are also
consistent with normal SNeIa at the spectroscopically determined
redshift of $z=1.3883$; however, the observed flux is a constant
factor of $\sim$30 times too bright in each band over time as compared
to the templates. At minimum, this shows that the peak luminosities
inferred from the light-curve widths of some SNeIa will deviate
significantly from the established, empirical relation used by
cosmologists. We argue on physical grounds that the observed fluxes do
not reflect an intrinsically luminous SNIa, but rather \afx\ is a
normal SNIa whose flux has been magnified by an external source. The
only known astrophysical source capable of such magnification is a
gravitational lens. Given the lack of obvious lens candidates, such as
galaxy clusters, in the vicinity, we further argue that the lens is a
supermassive black hole or a comparatively low-mass dark matter
halo. In this case, the lens continues to magnify the underlying host
galaxy light. If confirmed, this discovery could impact a broad range
of topics including cosmology, gamma-ray bursts, and dark matter
halos.

\end{abstract}


\section{Introduction}

Type\,Ia supernovae (SNeIa) represent a remarkably homogeneous class
of cataclysmic explosions. While some debate remains over the precise
physical nature of their progenitors and explosion mechanism
\citep[e.g.][]{howell2011}, a large body of evidence shows that SNeIa
reach nearly standard peak luminosities
\citep[e.g.][]{yasuda_fukugita2010}, and there is a strong correlation
between this and other observables, most notably their light-curve
shapes \citep{phillips1993}, which can be used to standardize this
peak power. This relation has famously been exploited to probe the
cosmology of our universe \citep[e.g.][]{riess1998, perlmutter1999}.

SNeIa are understood from theoretical work to be thermonuclear
explosions involving C/O white dwarf stars, a fact which is now
supported by observational constraints on the progenitor systems
\citep{nugent2011, bloom2012}. The peak luminosities of SNeIa and the
tight relationship between these and their light-curve shapes can be
understood in terms of the mass of radioactive $^{56}$Ni produced in
the first seconds of the explosion and the opacities of the ejected
material \citep[e.g.][]{kasen2007}.

Recently, \citealt{chornock2013} (hereafter C13) have presented a
transient, \afx, from the Panoramic Survey Telescope \& Rapid Response
System 1 (Pan-STARRS1; \citealt{kaiser2010}) survey with a light-curve
shape similar to SNeIa, but with an apparent luminosity far larger
($\sim 4.1 \times 10^{44}$\,erg\,s$^{-1}$). The redshift can be set
precisely at $z=1.3883$ based on narrow interstellar emission lines
presumably from the relatively faint, and compact galaxy within
$0\farcs1$ of the transient. The apparent luminosity of \afx\ is
similar to so-called superluminous supernovae \citep{galyam2012}, but
the colors and fast evolving light curve are not.

In this letter we argue that \afx\ is a normal SNIa with a normal peak
luminosity that has been magnified by a gravitational lens. We show
that the spectra are consistent with normal SNeIa in \S\ref{spectra}.
In in \S\ref{phot}, we show this classification is supported by the
photometry of \afx, but the fluxes are much brighter than the
light-curve width would suggest. In \S\ref{lens} we consider various
gravitational lensing scenarios to explain the anomalous brightness of
\afx. We conclude in \S\ref{conclusions} with a summary of our results
and discuss the implications of this discovery.

\section{Spectroscopic Classification}\label{spectra}

We first consider the spectroscopic nature of \afx, as spectroscopic
features define the classification system of supernovae
\citep[e.g.][]{wheeler_harkness1990, filippenko1997}. Our analysis is
limited to the four spectra presented by C13. We assign spectroscopic
classifications using the SN template $\chi^2$ fitting package, {\tt
  superfit} \citep{howell2005}. We employed 157 representative SNIa
templates with a roughly Gaussian distribution of phases ($7 \pm 13$
days), 135 SNIbc templates ($12 \pm 15$ days), and 95 SNII
templates. These include normal supernovae (e.g. 1994D, 1992A, 1994I,
1984L) as well as peculiar (e.g. 1991bg, 1991T, 1998bw). The templates
are redshifted, (de)reddened and combined with galaxy templates to fit
the observed spectra. Based on the host galaxy SED presented in C13,
we only consider a standard Sc type galaxy template below, although
our results do not change significantly if we adopt other templates.

\begin{figure}
\begin{center}
 \includegraphics[width=\linewidth]{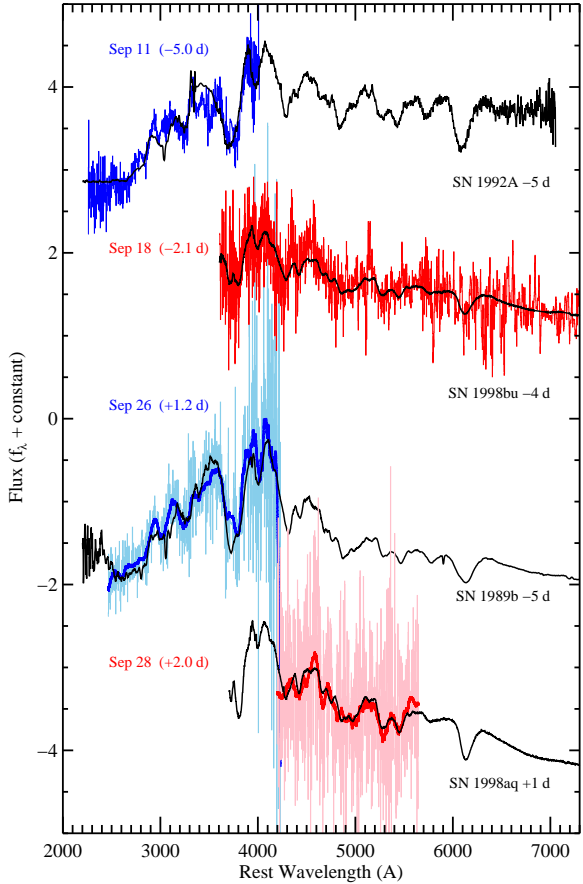} 
 \caption{Spectra of \afx\ (with contaminating galaxy light removed)
   compared to templates of normal SNIa as matched using {\tt
     superfit}. Phases for \afx\ are rest-frame days after the derived
   epoch of B-band maximum. The last two spectra are shown both before
   and after smoothing (light and dark colors, respectively).}
   \label{fig:spec}
\end{center}
\end{figure}

\begin{figure}
\begin{center}
 \includegraphics[width=\linewidth]{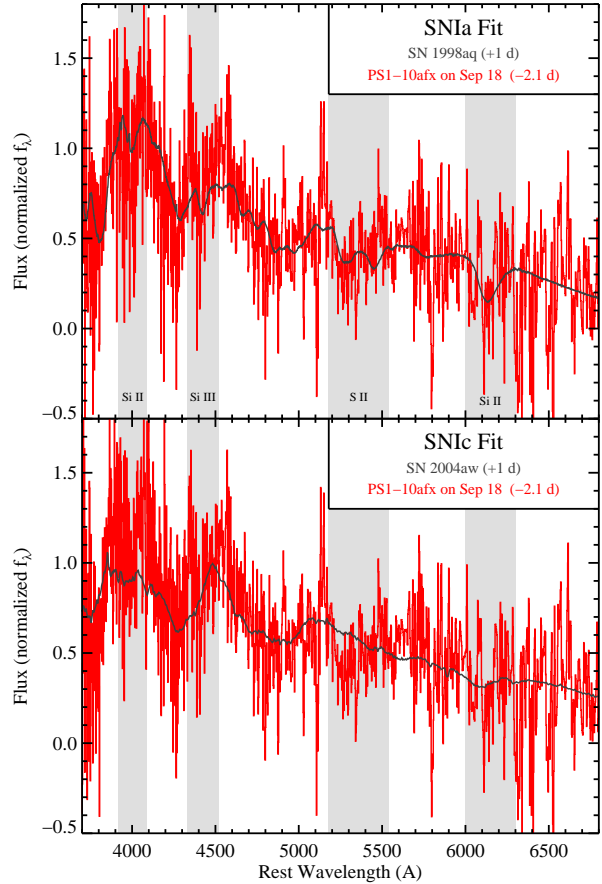} 
 \caption{Comparison of the near maximum light, rest-frame spectra of
   \afx\ to a normal SNIa (upper panel) and a representative SNIc
   (lower panel). Features demarcating the difference between the SNIa
   and SNIc types are highlighted in light gray. }
   \label{fig:spec2}
\end{center}
\end{figure}

The (observer-frame) NIR spectrum taken on 2010 September 18 covers
the rest-frame optical wavelength range over which the defining
spectroscopic signatures of supernovae can be found. Using {\tt
  superfit} we find that the spectrum is well matched by normal SNeIa
such as 1998aq and 1998bu near maximum light (see figures
\ref{fig:spec} and \ref{fig:spec2}). The top 5 matches have an average
phase relative to B-band maximum of $-3.2 \pm 3.7$\,days. SNIc
templates provide poorer quality fits and must be artificially
de-reddened by large amounts ($A_V \sim -2$). Likewise, subluminous,
SN\,1991bg-like SNeIa cannot match the features or relatively blue
continuum of \afx, although SN\,1986G also yields an acceptable fit
with $A_V = -2$. Matches to SNIa with more shallow silicon, such as
1999aa, are also allowed. Such fits favor less galaxy light
contamination, which can wash out the \ion{Si}{2} `6150' feature. The
\ion{Si}{2} `5800' feature falls in a strong telluric absorption band,
so we cannot compute the pseudo-equivalent width ratio to definitively
identify the SNIa subgroup to which \afx\ belongs.

Figure \ref{fig:spec2} emphasizes that several signatures of the
SNIa class are present in this near maximum light spectrum of \afx. In
addition to the \ion{Si}{2} features, which define this class, there
is a strong \ion{S}{2} ``W'', which is not seen in other supernova
types.

We also used the spectral cross-correlation package, {\tt snid}
version 5.0 with version 2.0 of the templates to classify the spectra
\citep{blondin2007}. Unlike {\tt superfit}, {\tt snid} is not able to
correct for contaminating galaxy light and it ignores the overall
continuum shape. The only acceptable matches (those with the
quality-of-fit parameter ${\rm rlap} > 5$) are for SNIa templates. The
best match is to the normal SNIa, 2008Z (${\rm rlap}=6.3$), but some
SN\,1991T-like templates match as well (e.g. SN\,1998es; ${\rm
  rlap}=5.1$). The best-fitting SN\,1991bg-like template is SN\,1999by
(${\rm rlap}=3.3$), and the best-fitting SNIc template is SN\,2004aw
(${\rm rlap}=3.1$).

Using {\tt superfit}, we similarly find that the other three spectra
are best fit by SNIa templates.  The (observer-frame) optical spectrum
obtained on 2010 September 26 and the (observer-frame) NIR spectrum
obtained two days later are noisier than the first spectra, but they
are each of sufficient quality to uniquely classify with {\tt
  superfit}. Both of these spectra are again consistent with normal
SNeIa. The fits show that below $\sim$2800\,\AA\ and above $\sim$6000
\AA, most of the signal is from contaminating galaxy light, which
would be consistent with the host SED fit in C13. Assigning this light
instead to the transient, C13 disfavor classifying \afx\ as a SNIa.

To summarize, any of the four spectra presented in C13 can be used to
classify \afx\ as a SNIa. The template fits set the redshift of the
supernova at $z=1.38$ (to a typical precision of about $\pm 0.01$)
independent of the host redshift. We thus securely conclude that
\afx\ can be spectroscopically classified as a SNIa at a redshift
consistent with its apparent host galaxy.

\section{Photometric Classification}\label{phot}

Spectroscopically normal SNeIa have similar light curves
\citep[e.g.][]{hicken2009a, ganeshalingam2010}. In this section, we
test whether \afx\ is photometrically consistent with normal
SNeIa using the photometry from C13.

To predict the expected observer-frame magnitudes for a normal SNIa at
a redshift of $z=1.3883$ in each of the photometric bands employed, we
use the spectral templates of \citet{hsiao2007} and a flat, $H_0 =
74$\,km\,s$^{-1}$, $\Omega_m = 0.27$ cosmology. The templates
represent the average from over a thousand individual observations of
normal SNeIa. We adjust the pre-maximum template phases to have a rise
time of 18 days, the average value for SNeIa with normal B-band
declines over 15 days after peak, $\Delta m_{15}(B) = 1.1$\,mag
\citep{ganeshalingam2011}.

\begin{figure}
\begin{center}
 \includegraphics[width=\linewidth]{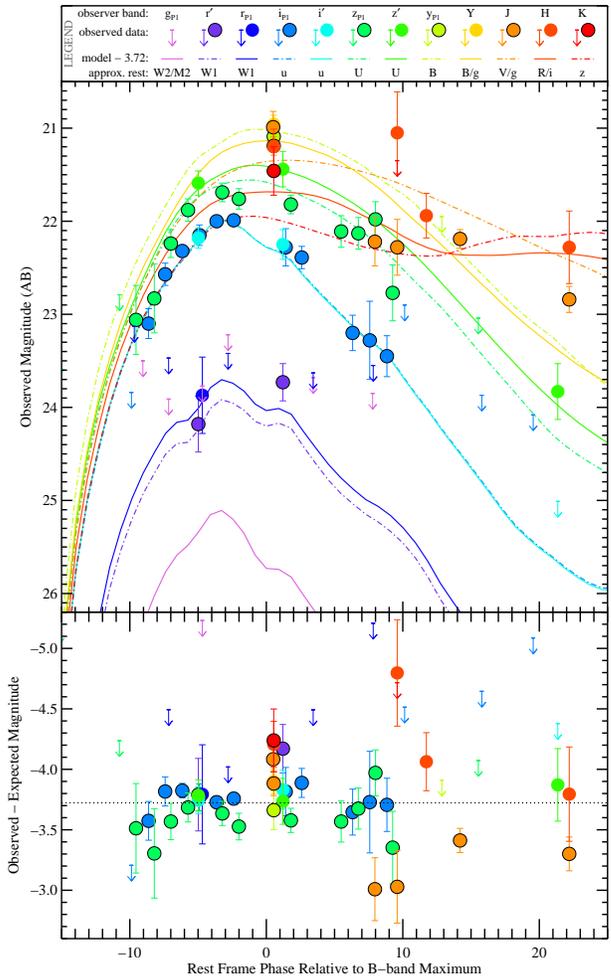} 
 \caption{Observed photometry of \afx\ from C13 compared to the best
   fit SNIa templates. The lower panel shows the observed minus
   expected magnitudes, with the template's peak brightness set by the
   $\Delta m_{15}(B)$ to peak-luminosity relation and our chosen
   cosmology. The dotted line is the best-fit magnitude bias, which is
   used to shift the template curves in the main panel to match the
   data.  }
   \label{fig:phot}
\end{center}
\end{figure}

Using {\tt mpfit.pro} in IDL, we fit the Hsiao templates to the
observed photometry (detections only) with three free parameters: the
epoch of B-band maximum, a single ``stretch'' term, and a global flux
scaling term. We first consider only the photometry from C13 measured
after removal of quiescent light using PSF-matched templates taken
with the same instruments and filters. These data include 24
detections in the \rps, \ips, \zps, and \yps\ bands over 19 rest-frame
days. The best fit template has a $\chi^2$/DoF of $20.4 / 20 = 1.02$
and indicates B-band maximum occurred on ${\rm MJD} = 55462.1 \pm 0.6$
(see Fig. \ref{fig:phot}). From the stretched template we measure
$\Delta m_{15}(B) = 1.22 \pm 0.09$\,mag. This rules out SNeIa like
1999aa, which have $\sim$0.8\,mag declines. For $H_0=74$, the MLCS2k2
light-curve fitting package \citep{jha2007} predicts such SNeIa
should have peak absolute magnitudes of $M_{B} = -19.08 \pm 0.1$. The
observed magnitudes are $3.72 \pm 0.03$\,mag brighter. C13 note that
\afx\ is 3.8 magnitudes brighter than HST04Sas, a spectroscopically
confirmed SNIa at $z=1.39$ \citep{riess2007}.

The Hsiao template fit also suggests a rise to B-band maximum of
$\sim$16 days, which is normal for SNeIa
\citep[e.g.][]{ganeshalingam2011}; however, prior to the first
detection, there is a \ips\ (rest-frame 3160\,\AA) limit 0.5\,mag
below the template prediction. We attribute this to a problem in the
template. \citet{brown2012} have shown that the UV light curves of
SNeIa do not rise as smoothly as the templates assume. Rather, there
is a rapid rise in flux near the epoch in question. The Hsiao
templates over predict the UV flux at earlier
phases. \citet{brown2012} demonstrate that fitting a ``fireball'' ($L
\propto t^2$) model to UV observations of SN\,2011fe near and after
this phase will underestimate the true rise time; this would explain
the faster rise favored by C13.

The remaining photometry from C13, based on cross instrument
measurements, is also shown in figure \ref{fig:phot}. C13 point out
that their $z\arcmin$ measurements near maximum light are 0.2\,mag
brighter than their \zps\ data from similar phases, but the $r\arcmin$
and $i\arcmin$ measurements appear consistent with the \rps\ and
\ips\ bands, respectively. They do not, however, discuss systematic
offsets between the four different instruments employed for the NIR
photometry, or how cross filter differences may vary with time as the
spectra of \afx\ evolve (i.e. the $0.2$\,mag offset between the $z$
bands is expected for a SNIa near maximum, but the difference
increases to $\sim$0.4\,mag on day $+23$). Some of the NIR
measurements have the host light removed not through PSF-matched image
subtraction, but numerically. It is not clear if the supernova light
was measured in the same 1.7\arcsec\ radius aperture as the host
photometry, or how C13 may have accounted for systematic differences
between the measurements, but some of the J-band measurements of
\afx\ have errors in flux that are smaller than the host's
error. Including the NIR measurements, the $\chi^2$/DoF for the
template fit increases to 2.0, but allowing for a small systematic
error ($\sim$0.1\,mag) results in a good fit ($\chi^2$/DoF = 1.2).

From the overall excellent agreement in the spectra, light-curve
shapes, and colors predicted from the SNIa templates, we conclude that
\afx\ is a normal SNIa. The well established physical models for SNeIa
suggest that the observed flux excess is not intrinsic to the
supernova itself; rather, \afx\ has been magnified by an external
source.

\section{Source of the Magnification}\label{lens}

As a normal SNIa, we can calculate the actual luminosity of \afx\ and
compare this to its inferred luminosity to derive the signal
magnification. This results in a magnification factor of
$30.8^{+5.6}_{-4.8}$ assuming a SNeIa dispersion of $\sigma=0.18$\,mag
after $\Delta m_{15}(B)$ correction. The only astrophysical object
capable of such large, achromatic signal magnification is a
gravitational lens. In this section, we consider the possible lensing
scenarios that may explain the anomalous properties of \afx.

Massive galaxy clusters are perhaps the best known sources of strong
gravitational lensing. C13 point out that there is no evidence for
such a cluster in the Pan-STARRS1 data. A red sequence cluster finder
\citep{finoguenov2010} also shows no evidence for a massive cluster in
the deeper CFHT images \citep{gwyn2008}. With a galaxy cluster lens,
we would also expect strong distortion of the host into an arc (or
arcs). We thus conclude that \afx\ is not lensed by a galaxy cluster.

Moving to small scales, we next consider micro-lensing events. It is
possible for a massive source in our Galaxy to greatly magnify light
from a background source, but the peculiar motion of the lens across
the plane of the sky requires the magnification factor to vary with
time. The magnification of typical micro-lensing events varies by a
factor of a few over $\sim$20 day periods \citep[e.g.][]{besla2013},
whereas \afx\ appears consistent with a constant magnification over at
least 2 months.

Individual galaxies may magnify supernovae as well
\citep[see][]{mortsell2001, oguri_marshall2010}. There is a galaxy to
the south east with $z_{\rm photo} \sim 0.6$ \citep{ilbert2006}, but
the separation, about 8\arcsec, is far too large for this to be a
viable lensing candidate. Alternatively, the lensing galaxy could be
directly in front of, and thus confused with the host. In this case,
the lensing galaxy must be faint or physically compact to be
compatible with the available imaging. The lensing galaxy also cannot
have much dust nor can it harbor an optically luminous AGN as the
photometry of \afx\ do not indicate reddening, and the spectra do not
show emission lines aside from the nebular lines in the SN
rest-frame. These constraints may allow for a compact red galaxy as
the lens, and the light from this object may dominate the true host
galaxy's NIR emission.

\begin{figure}
\begin{center}
 \includegraphics[width=\linewidth]{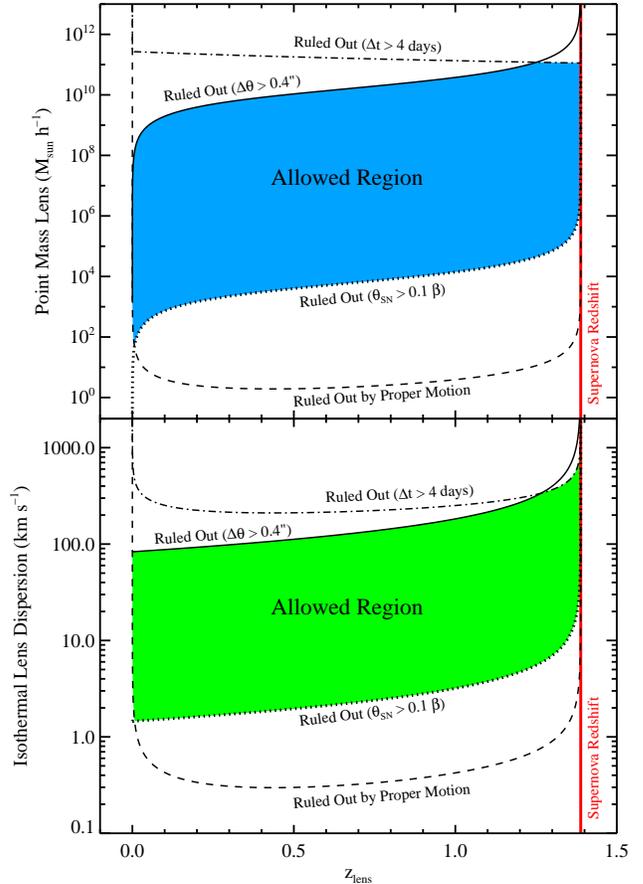} 
 \caption{Plausible gravitational lensing scenarios. {\it Top panel:}
   Mass-lensing redshift ($z_{\rm lens}$) plane for a point mass. {\it
     Lower panel:} Velocity dispersion-lensing redshift plane for an
   isothermal lens.}
   \label{fig:lens}
\end{center}
\end{figure}

The apparent lack of obvious, luminous material at the lens redshift
could suggest the lens is a dark matter halo with little baryonic
material, such as the compact red galaxy noted above, or an isolated,
supermassive black hole. Figure \ref{fig:lens} depicts plausible lens
parameters for a point mass and for an isothermal body that meet the
following criteria.

We require a large magnification, so the supernova can only be
slightly misaligned from the lensing object with an Einstein radius,
$b$, by an angle, $\beta$. For a point mass we have
\begin{equation}
b=\left[\frac{4GM}{c^2} \frac{\Dls}{\Ds\Dl}\right]^{1/2}\,,
\label{einst_rad}
\end{equation}
where $\Dls$, $\Dl$ and $\Ds$ are the angular diameter distance
between the lens and the source, between us and the lens and between
us and the source, respectively. For an isothermal body,
\begin{equation}
b=4\pi\,\left(\frac{\sigma}{c}\right)^2 \frac{\Dls}{\Ds}\,,
\label{einst_rad_sis}
\vspace{0.1cm}
\end{equation}
where $\sigma$ is the velocity dispersion. The angular separation
between the two images is given by $\Delta \theta \simeq 2 b$.
\afx\ is apparently unresolved in the Pan-STARRS1 images, so we adopt
$\Delta\theta < 0.4$\arcsec. For the point mass, the sum of the
magnifications of the two images is given by
\begin{equation}
\mu_{\rm tot} = \frac{(2+x^2)}{x(x^2+4)^{1/2}}\,,
\label{summag}
\end{equation}
with $x = \beta / b$. Setting $\mu_{\rm tot} = 30.8$, we find that $x
= 0.03$. For the isothermal body,
\begin{equation}
\mu_{\rm tot} = \frac{2b}{\beta}=\frac{2}{x}\,,
\label{summag_sis}
\end{equation}
and thus $x = 0.06$.

The angular size of \afx's photosphere must be smaller than the
misalignment angle (i.e. $\theta_{\rm SN} < 0.1 \beta$) to avoid
changes in the effective magnification over time. Also, the angle by
which the lens moves with respect to \afx\ during the two months of
observations cannot be large or, like the micro-lensing examples cited
above, the magnification would vary. These constraints are also
depicted in figure \ref{fig:lens}. In addition, the light curve fit
from \S\ref{phot} suggests that the time delay, $\Delta t$, between
images is small (we further note a lack of evidence for mircolensing
distortions; \citealt{dobler_keeton2006}).

\section{Conclusions and Discussion}\label{conclusions}

\afx\ is a Type\,Ia supernova. This is strictly true since the spectra
of \afx\ show strong \ion{Si}{2} features and no hydrogen, which are
the defining characteristics of the SNIa class. In addition, normal
SNeIa can explain the light-curve shape and broad band colors, as well
as the phase of the spectra with respect to the photometric B-band
maximum light epoch while other varieties of supernovae are
incompatible with these observations. However, we find that the flux
measurements for \afx\ are systematically $\sim$30 times brighter than
can be explained by a SNIa with the observed light-curve shape. At a
minimum, this shows that some SNeIa, which otherwise appear perfectly
normal, can have fluxes greatly exceeding the well known luminosity to
light-curve width relation, which is fundamental to the use of SNeIa
as cosmological probes.

We next consider the physical origin of \afx's apparently anomalous
flux. We argue that this cannot be representative of the intrinsic
luminosity of the supernova based on physical grounds. The spectra of
SNeIa have been extensively modeled and are found to be the results
of thermonuclear explosions consuming carbon and oxygen to produce
intermediate mass elements and iron peak elements including, most
importantly, $^{56}$Ni, which radioactively decays to provide the input
power for the light curve. C13 have already pointed out that $^{56}$Ni
cannot produce both a luminosity of a few$\times
10^{44}$\,erg\,s$^{-1}$ as well as the 2-week time scale rise and fall
of \afx: the mass of $^{56}$Ni required would exceed the allowed
ejecta mass, which largely determines the light-curve time
scale. \afx\ can thus not have both the spectroscopic features
observed, which imply a thermonuclear origin, and an intrinsically
high peak luminosity, which would require more complete thermonuclear
burning than the light curves can accommodate. We conclude that
\afx\ is a normal SNIa with a normal luminosity and that the
anomalously bright flux observed is a result of factors external to
the supernova.

The only known astrophysical scenario for achromatically magnifying
the flux of an object is a gravitational lens. We have presented the
physical constraints on a point mass lens and an isothermal lens based
on geometrical considerations and the observed magnification
factor. It should be noted that this magnification is independent of
the existence of the supernova; it applied and still applies to the
host environment of \afx.

This prediction of a persistent lensing source provides a test of our
hypothesis. High spatial resolution images from {\it HST} may be able
to resolve the Einstein ring from the magnified host galaxy. If the
lens is a compact red galaxy then color information should distinguish
it from the blue, star-forming galaxy in the background. If a
foreground galaxy is not detected, then this would either suggest that
more exotic lensing systems, such as free floating black holes, are
required or it could support the hypothesis of C13 that a new class of
superluminous supernovae is required to explain \afx.

New physics may not be needed, however, since known physical
processes can already provide a satisfying explanation.
\citet{oguri_marshall2010} have predicted that $\sim$0.1 SNeIa
strongly lensed into resolved pairs or quads should be found in the
Pan-STARRS1 Medium Deep Survey. Revising this calculation to include
SNeIa with unresolved images, we calculate 0.8 events are expected. It
is thus statistically plausible for Pan-STARRS1 to have detected a
gravitationally lensed SNIa as we have concluded. Due to Malmquist
bias in the selection of candidates, objects with larger
magnifications are more likely to be detected and followed up. Monte
Carlo simulations (following \citealt{oguri_marshall2010}) show that
magnification factors of $\sim$20 are expected from strong lenses in
the flux-limited Pan-STARRS1 sample even though there are many more
SNeIa with lower magnifications in a given volume. Such highly lensed
SNeIa may impose a larger systematic error in future, high precision
cosmological measurements than previously thought
\citep[e.g.][]{holz_linder2005}.

A larger ``milli-lensing'' sample may give some insight into the
distribution of these lensing systems. For example, such observations
can be used to obtain constraints on the number of dark subhalos
around galaxies that have been speculated to exist
\citep{simon2007}. Sources at larger redshifts (e.g. $z > 1$) have a
substantially larger optical depth to foreground lenses and are thus
much more likely to be lensed. We speculate that other flux limited
samples of transient objects may already be contaminated with lensed
events. In particular, some of the massive, red galaxies coincident
with ``dark'' gamma-ray bursts may be foreground lenses to even higher
redshift events \citep[][]{perley2013}.

It may become increasingly common to find lensed transients like
\afx\ as next generation optical transient surveys with the Dark
Energy Camera, the Hyper-Suprime Camera, and the Large Synoptic Survey
Telescope, begin deep, wide-area surveys. Such discoveries may be
exploited to probe the expansion of our universe
\citep[e.g.][]{oguri_kawano2003,linder2011}, or as tests of gravity
\citep[e.g.][]{smith2009}.

\acknowledgments We thank Kevin Bundy and Christopher Kochanek for
comments. This work was supported in part by the Kakenhi Grant-in-Aid
for Young Scientists (B)(24740118) from the Japan Society for the
Promotion of Science, the World Premier International Research Center
Initiative, MEXT, Japan, and the FIRST program, ``Subaru Measurements
of Images and Redshifts (SuMIRe)''.

\bibliographystyle{aa}

\end{document}